\documentclass[reprint,pra,twocolumn,showpacs,superscriptaddress,aps,longbibliography]{revtex4-2}

\usepackage{graphicx}
\usepackage{epstopdf}

\usepackage[T1]{fontenc}
\usepackage[applemac]{inputenc}
\usepackage{lmodern}
\usepackage[english]{babel}

\usepackage{ae}
\usepackage{siunitx}

\usepackage{amsmath,amssymb,natbib,bm}
\usepackage{psfrag}
\usepackage{subfigure}
\usepackage{amsthm}
\usepackage{algorithm}
\usepackage{algpseudocode}

\usepackage[colorlinks]{hyperref}
\hypersetup{%
        plainpages=true,
        breaklinks=true,
        hypertexnames=false,
        pageanchor=true,
        colorlinks=true,
        linkcolor={blue},
        citecolor={blue},
        urlcolor={blue},
        anchorcolor={black}
      }

\usepackage{mleftright} 

\newcommand{\figref}[1]{\mbox{Fig.~\ref{#1}}}

\newcommand{\secref}[1]{\mbox{Sec.~\ref{#1}}}
\newcommand{\appref}[1]{\mbox{Appendix~\ref{#1}}}
\renewcommand{\eqref}[1]{\mbox{Eq.~(\ref{#1})}}

\newcommand{\figpanel}[2]{Fig.~\hyperref[#1]{\ref*{#1}(#2)}}
\newcommand{\figpanels}[3]{Fig.~\hyperref[#1]{\ref*{#1}(#2)--(#3)}}
\newcommand{\figpanelNoPrefix}[2]{\hyperref[#1]{\ref*{#1}(#2)}}

\newcommand{\bra}[1]{\langle #1|}
\newcommand{\ket}[1]{|#1\rangle}

\newcommand{\be}{\begin{equation}}
\newcommand{\ee}{\end{equation}}
\newcommand{\bea}{\begin{eqnarray}}
\newcommand{\eea}{\end{eqnarray}}

\begin{document}

\author{Akshay Gaikwad}
\email{akshayga@chalmers.se}
\affiliation{Department of Microtechnology and Nanoscience, Chalmers University of Technology, 41296 Gothenburg, Sweden}


\author{Manuel Sebastian Torres}
\affiliation{Department of Microtechnology and Nanoscience, Chalmers University of Technology, 41296 Gothenburg, Sweden}


\author{Anton Frisk Kockum}
\email{anton.frisk.kockum@chalmers.se}
\affiliation{Department of Microtechnology and Nanoscience, Chalmers University of Technology, 41296 Gothenburg, Sweden}

\title{Quantum measurement tomography with mini-batch stochastic gradient descent}

\begin{abstract}

Drawing inspiration from gradient-descent methods developed for data processing in quantum state tomography [\href{https://iopscience.iop.org/article/10.1088/2058-9565/ae0baa}{Quantum Sci.~Technol.~\textbf{10}, 045055 (2025)}] and quantum process tomography [\href{https://journals.aps.org/prl/abstract/10.1103/PhysRevLett.130.150402}{Phys.~Rev.~Lett.~\textbf{130}, 150402 (2023)}], we introduce stochastic gradient descent (SGD) algorithms for fast quantum measurement tomography (QMT), applicable to both discrete- and continuous-variable quantum systems---thus completing the tomography trio. A measurement device or detector in a quantum experiment is characterized by a set of positive operator-valued measure (POVM) elements; the goal of QMT is to estimate these operators from experimental data. To ensure physically valid (positive and complete) POVM reconstructions, we propose two distinct parameterization schemes within the SGD framework: one leveraging optimization on a Stiefel manifold and one based on Hermitian operator normalization via eigenvalue scaling. Within the SGD-QMT framework, we further investigate two loss functions: mean squared error, equivalent to L2 or Euclidean norm, and average negative log-likelihood, inspired by maximum likelihood estimation. We benchmark performance against state-of-the-art constrained convex optimization methods. Numerical simulations demonstrate that, compared to standard methods, our SGD-QMT algorithms offer significantly lower computational cost, superior reconstruction fidelity, and enhanced robustness to noise. We make a Python implementation of the SGD-QMT algorithms publicly available at \href{https://github.com/agtomo/SGD-QMT}{github.com/agtomo/SGD-QMT}.

\end{abstract}

\date{\today}

\maketitle


\section{Introduction}


In recent years, the field of quantum information science has witnessed substantial progress in the development of quantum characterization, verification, and validation (QCVV) protocols~\cite{hashim-prxquantum-2025, eisert-nrp-2020, proctor-nrp-2025, roth-prxquant-2021, kenneth-prapplied-2022, nielsen-njp-2021, kenneth-prxquant-2021, roth-prxquant-2023, hashim-npj-quantinfo-2023, gambetta-pra-2018, zeng-prr-2020, gambetta-pra-2019, nielsen-quantum-2021, flammia-njp-2014, gambetta-prl-2012, proctor-prl-2022, helsen-npj-quant-info-2019, simon-ieeetrans-2021}.
These protocols encompass a wide spectrum of techniques, generally organized into four major categories~\cite{hashim-prxquantum-2025}: experimental device characterization, tomographic methods, randomized benchmarking, and task-specific benchmarking.
Collectively, they provide essential tools for ensuring the reliable and scalable operation of quantum hardware. In particular, QCVV protocols not only serve as benchmarks for assessing device performance but also as diagnostics for identifying and quantifying error mechanisms, and as practical guides for optimizing control, calibration, and error-mitigation strategies, thereby laying the foundation for advancing from noisy intermediate-scale quantum (NISQ) systems toward scalable, fault-tolerant quantum computation~\cite{preskill-quantum-2018, gebhart-nrp-2023, elbennrp-2023, arute-nature-2019, zhong-science-2020}.
The continued refinement of QCVV techniques is thus critical for advancing quantum technologies from proof-of-principle demonstrations toward robust, large-scale implementations.

\begin{figure}[t]
    \centering
    \includegraphics[width=0.8\linewidth]{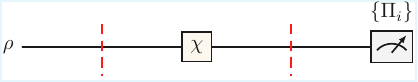}
    \caption{Quantum circuit illustrating the three fundamental components of a quantum experiment: the quantum state, described by the density matrix $\rho$; the quantum process, represented by the process matrix $\chi$; and the measurement apparatus, characterized by a set of POVM elements $\{ \Pi_i \}$.}
    \label{fig:qc}
\end{figure}

When applying QCVV techniques to quantum processors, the operation of the processor can, in general, be understood in terms of three fundamental components~\cite{chuang-2010, divincenzo-2000, georgescu-nrp-2020, faist-prl-2016}: quantum states (inputs), quantum processes or gates (operations), and measurement readouts or detection (outputs); see \figref{fig:qc}. Each component in this trio plays a vital role in the functionality of the processor, and errors or imperfections in any part can lead to incorrect or unreliable computational outcomes. Therefore, it is extremely important to have efficient strategies for comprehensive characterization of the whole trio. 

Among QCVV tools for comprehensive characterization of quantum systems, tomographic techniques are some of the most widely used. These techniques include: (i) \textit{quantum state tomography} (QST)~\cite{gale-pra-1968, shang-pra-2017, gross-prl-2010, Qi-npj-quantu-info-2017, Shahnawaz-prl-2021}, which reconstructs the density matrix $\rho$ describing an unknown quantum state; (ii) \textit{quantum process tomography} (QPT)~\cite{chuang-jmo-09, white-prl-2004, kim-natcom-2017, rod-prb-2014, gaikwad-pra-2018, gaikwad-sci-rep-2022, knee-pra-2018}, which estimates the action of a quantum process through a process matrix $\chi$; and (iii) \textit{quantum measurement tomography} (QMT)~\cite{jaromir-pra-2001, lundeen-natphys-2009, ilia-scipost-2021, zoltan-prr-2023, xiao-automatica-2023, julia-prr-2025}, also referred to as \textit{quantum detector tomography}, which characterizes the measurement apparatus modeled by a set of positive operator-valued measure (POVM) elements $\{ \Pi_i \}$. All these three tomographic techniques are statistical procedures that rely on experimental data and an estimator---an algorithm that maps the collected data to an estimate of the underlying state, process, or detector. In this article, we put forward and benchmark algorithms for such data processing in the case of QMT, inspired by earlier work on that topic for QST~\cite{gaikwad-arxiv-2025} and QPT~\cite{ahmed-prl-2023}.

While QST, QPT, and QMT share similar conceptual frameworks and workflows, each is governed by distinct physical constraints~\cite{hashim-prxquantum-2025, roth-prxquant-2023}:
\begin{align}
&\text{QST:} \hspace{0.1cm} \rho \in \mathbb{C}^{d \times d} \hspace{0.1cm} \text{s.t.} \hspace{0.1cm} \rho = \rho^\dag, \hspace{0.1cm} {\rm Tr}(\rho) = 1 , \hspace{0.1cm} \& \hspace{0.1cm} \rho \geq 0 ,
\label{qs}
\\
&\text{QPT:} \hspace{0.1cm} \chi \in \mathbb{C}^{d^2 \times d^2} \hspace{0.1cm} \text{s.t.} \hspace{0.1cm} \chi \geq 0 \hspace{0.1cm} \& \hspace{0.1cm} \sum_{m,n} \chi_{mn} E_m^\dag E_n = I ,
\label{qp}
\\
&\text{QMT:} \hspace{0.1cm} \mleft\{ \Pi_i \in \mathbb{C}^{d \times d} \mright\}_{i=1}^k \hspace{0.1cm} \text{s.t.} \hspace{0.1cm} \sum_{i=1}^k \Pi_i = I \hspace{0.1cm} \& \hspace{0.1cm} \Pi_i \geq 0 \hspace{0.05cm} \forall i,
\label{qd}
\end{align}
where $d$ is the dimension of the Hilbert space, $\{ E_i \}$ and $I$ in \eqref{qp} are a fixed operator basis set and the $d \times d$ dimensional identity operator, respectively, and the scalar $k$ in \eqref{qd} denotes the number of POVM elements used to model the given measurement apparatus.

There is a wide range of QST and QPT methods that effectively incorporate their respective physical constraints. In contrast, significantly fewer techniques have been proposed and studied for QMT. While many QST algorithms can be readily extended to QPT through the Choi-Jamiolkowski isomorphism---also known as the \textit{state-channel duality} theorem~\cite{choi-laa-1975, leung-2003, jiang-pra-2013}---their extension to QMT becomes non-trivial due to the distinct constraints governing QMT.

Moreover, like QST and QPT, QMT is also hindered by the exponential scaling of required measurement data and the computational burden of post-processing. In fact, QMT poses an even greater challenge than QST in some sense, as it necessitates the reconstruction of an entire POVM set $\{ \Pi_i \}$ rather than a single density matrix. The cardinality of this POVM set typically scales linearly with the Hilbert-space dimension (exponentially with qubit number); for instance, projective measurements in a given Pauli basis on an $N$-qubit system involve estimating $2^N$ projectors, each of dimension $2^N \times 2^N$. Consequently, there is a critical need for QMT data-processing algorithms that are computationally efficient, adaptable to various optimization frameworks, and capable of enforcing the physical constraints in \eqref{qd}.


Among the relatively few methods for QMT, most approaches centered around convex optimization techniques, particularly based on semi-definite programming. A common strategy, used in recent works~\cite{julia-prr-2025, zambrano-arxiv-2025}, involves first applying a linear inversion estimator to obtain an initial, unconstrained (invalid) estimate of the POVM elements, followed by a convex projection step to enforce physical constraints such as positivity and completeness. 
In this approach, the quality of the POVM elements reconstructed via constrained convex optimization (CCO) is fundamentally limited by the unconstrained estimates obtained through linear inversion. It is well known that the linear inversion problem often is \textit{ill-conditioned}, due to the high condition number of the sensing matrix~\cite{hinham-book-2002, kahan-1966}. As a result, even small amounts of noise in the data can lead to large errors in the unconstrained estimates~\cite{nori-pra-2014}, which in turn degrade the accuracy of the subsequent CCO refinement.

An alternative CCO framework for QMT employs semi-definite programming~\cite{zoltan-prr-2023}, where the optimization is performed directly on the measurement data. In this formulation, both the probe input states and the POVM elements are jointly optimized (up to gauge freedom) while CCO enforces the physical constraints required for QST and QMT. This yields a form of self-consistent QMT that is inherently robust to imperfections in probe-state preparation. Despite its conceptual elegance, the method is computationally intensive and has thus far been demonstrated numerically only for rank-1 Pauli projective measurements involving up to three qubits. Another recent work~\cite{Schapeler-qst-2025}, based on a similar CCO approach, uses classical high-performance computing hardware to demonstrate large-scale quantum photonic detector tomography.

These CCO-based methods leverage the CVX optimization package in Python, which provides a high-level interface to express convex optimization problems and integrates well with efficient back-end solvers (e.g., SCS and MOSEK)~\cite{diamond2016cvxpy, cvx-solvers}. While these CCO-based techniques are reliable and theoretically well-founded, they can be computationally intensive, particularly for high-dimensional systems, since the number and size of the POVM elements scale exponentially with system size.

Another approach to data processing for QMT is iterative maximum likelihood estimation (iMLE)~\cite{jaromir-pra-2001, Brida-njp-2012, chen-pra-2019}, where the POVM elements are iteratively updated to maximize the likelihood of the observed measurement data under a given noise model. However, the standard update procedure in iMLE often enforces the completeness constraint strictly throughout the optimization, but relaxes the positivity constraint during intermediate updates. As a result, one of the POVM elements may temporarily fall outside the physical (positive semidefinite) domain, 
requiring post hoc projection or regularization to restore validity---potentially impacting convergence behavior and estimation accuracy.

Here, inspired by the framework of data processing using stochastic gradient descent (SGD) for QST~\cite{gaikwad-arxiv-2025} and QPT~\cite{ahmed-prl-2023}, we introduce SGD-QMT. This completes the trio of gradient-descent based tomographic methods, offering an efficient framework for characterizing all three fundamental components of a quantum processor. Our methods can be applied to both discrete-variable (DV) and continuous-variable (CV) quantum systems.

We reformulate QMT as an iterative function minimization problem, utilizing \textit{mini-batch} SGD methods to efficiently handle large data sets. To ensure that the optimization remains confined within the space of physically valid POVM elements at all times, we propose two distinct parameterizations: (i) SM parameterization, which employs Stiefel-manifold optimization to preserve orthonormality (ensuring completeness) and positivity, and (ii) HONEST parameterization, which utilizes Hermitian operator normalization via an eigenvalue scaling technique to enforce physical validity. Additionally, we investigate two loss functions: (i) mean squared error (MSE) and (ii) average negative log-likelihood, inspired by MLE.

We assess the performance of our SGD-QMT algorithms by comparing them against established CVX-based convex optimization techniques. Our evaluation considers reconstruction quality, noise robustness, and computational time. For DV quantum systems, we conduct simulations up to six qubits, achieving convergence within a few seconds up to minutes on a standard laptop (18 GB of RAM, no dedicated GPU). We also demonstrate SGD-QMT on CV quantum systems, particularly focusing on photon detection (two-outcome measurement process) and photon-counting scenarios, where the number of POVM elements scales linearly with the Hilbert-space dimension. Numerical results show that our SGD-QMT methods consistently outperform traditional approaches across most of these metrics, with the HONEST parameterization combined with the MLE loss function proving to be the most effective overall (in particular for projective measurements).

These results highlight the performance and versatility of SGD-QMT methods across diverse quantum experiments. We therefore expect that SGD-QMT can become an important part of the QCVV toolbox for experimental applications in both development of quantum technologies and fundamental quantum information science. To help experimentalists and to support adoption and further research, we provide open access to our Python implementation of SGD-QMT at \href{https://github.com/agtomo/SGD-QMT}{github.com/agtomo/SGD-QMT}.

This article is structured as follows. In \secref{sec:Methods}, we first give an overview of the standard formalism for QMT, then present a detailed description of our SGD-QMT methods based on the SM and HONEST parameterizations, briefly review the state-of-the-art CCO methods we use for comparative analysis against our SGD-QMT algorithms,  and describe the measurement processes (which act as data sets for QMT methods) for both DV and CV systems that we used for benchmarking. In Sec.~\ref{sec:Results}, we present comprehensive numerical results and analysis. We conclude in Sec.~\ref{sec:conclusion} with a summary and a discussion on future directions. In the appendixes, we provide additional details on SGD-QMT for noisy data and on parameter-update rules and hyperparameters for our SGD-QMT algorithms.


\section{Methods}
\label{sec:Methods}

In this section, we present our SGD-QMT algorithms in detail and also provide an overview of state-of-the-art QMT data-processing methods for comparison. First, in \secref{sec:povms}, we review the POVM formalism and the standard linear-inversion approach used in QMT. Then, in \secref{sec:sgd-qmt}, we describe the formalism of our SGD-QMT based on two distinct parameterizations: the Stiefel manifold (SM) and the Hermitian operator normalization via eigenvalue scaling technique (HONEST). In Sec.~\ref{sec:cco}, we briefly review CCO methods, against which we benchmark our SGD-QMT algorithms, and in \secref{sec:customization}, we discuss some customization of these algorithms that become possible when one has some prior knowledge about the system. Finally, in Sec.~\ref{sec:datasets}, we outline the data sets (i.e., types of measurement processes or detectors) on which we implement QMT algorithms and define the performance metrics we employ to evaluate and compare the different QMT methods.

Note that, throughout this paper, we mainly use the terms \textit{measurement}, \textit{measurement device}, or \textit{measurement apparatus}---rather than \textit{quantum detector}---to align with the abbreviation QMT and maintain clarity and generality. Although these terms are conceptually equivalent, \textit{quantum detector} is more commonly associated with optical systems and is less prevalent in broader contexts.


\subsection{Standard formalism for quantum measurement tomography}
\label{sec:povms}

In quantum theory, a general $k$-outcome measurement (or detection) process on a $d$-dimensional quantum system is mathematically described by a set of $k$ $d \times d$-dimensional POVM elements ${\Pi} = \{ \Pi_1, \Pi_2, \ldots, \Pi_k \}$, also known as \textit{effects}~\cite{chuang-2010}. These operators satisfy the positivity ($\Pi_i \geq 0 \: \forall i$) and completeness ($\sum_{i=1}^k \Pi_i = I$) conditions.  According to the fourth postulate of quantum mechanics (also known as the \textit{Born rule}), the probability of obtaining the measurement outcome associated with $\Pi_i$ (equivalent to observing the $i$th effect), when measuring a quantum state $\rho_j$, is given by
\begin{equation}
p_{ij} = \text{Tr}(\Pi_i \rho_j ).
\end{equation}
The physical constraints defined for POVM operators ensure probabilities are non-negative ($p_{ij} \geq 0 \: \forall i,j$) and normalized ($\sum_{i} p_{ij}=1 \: \forall j$). 

The main objective of QMT is to estimate the set of unknown POVM operators $\{ \Pi_i\}_{i=1}^k$ that describe the measurement outcomes produced by an underlying uncharacterized measurement apparatus.  This objective is commonly achieved by probing the measuring device with an \textit{informationally complete} (IC) set of known input states $\{\rho_j \}$, whose cardinality typically scales quadratically with the dimension of the Hilbert space. Once the probabilities $\{ p_{ij} \}$ are obtained from experiments, one can formulate a system of linear equations of the form~\cite{julia-prr-2025}
\begin{equation}
\mathcal{A} \mathcal{X}_{\{\Pi_i\}} = P \quad \Rightarrow \quad \mathcal{X}_{\{\Pi_i\}} = \mathcal{A}^{-1} P ,
\label{eq:linear-inv}
\end{equation}
where $\mathcal{A}$ is the coefficient matrix (also called \textit{sensing matrix}). This matrix depends only on the chosen probe input states and the operator bases in which the POVM elements are represented, and can be computed analytically. The vector $\mathcal{X}_{\{\Pi_i\}}$ encodes the unknown parameters of the POVM elements to be reconstructed and $P$ is a flattened matrix that contains the experimentally measured probabilities $\{p_{ij}\}$. 

By directly solving \eqref{eq:linear-inv}, one can obtain the underlying set of POVM elements. However, linear inversion-based QMT not only produces unphysical results, as the reconstructed POVMs may violate the required positivity constraint, but also suffers from the fact that constructing and inverting  the coefficient matrix $\mathcal{A}$ becomes computationally hard as the system dimension increases. Moreover, the linear-inversion method is often \textit{ill-conditioned}, with the condition number of the coefficient matrix $\mathcal{A}$ exceeding 1~\cite{xiao-automatica-2023}. In such cases, even small errors in the data vector $P$ can lead to large errors in the solution $\mathcal{X}_{\{\Pi_i\}}$, making the linear-inversion estimation less robust to experimental noise. 


\subsection{Data processing with stochastic gradient descent for quantum measurement tomography}
\label{sec:sgd-qmt}

We now reformulate the standard linear-inversion-based QMT problem, described in \eqref{eq:linear-inv}, as an iterative optimization task using mini-batch stochastic gradient descent (SGD). We first discuss the loss functions and optimization strategies we use for this task, and then detail the parameterizations we use to ensure physical validity of the estimated POVMs throughout the optimization.


\subsubsection{Loss functions}
\label{sec:LossFunctions}

We consider two types of loss functions that can be directly computed from experimentally acquired data: (i) the \textit{mean squared error} (MSE)~\cite{xiao-automatica-2023}, which effectively minimizes discrepancy between experimentally observed and predicted outcomes (equivalent to the L2 norm) and (ii) the average of the negative logarithm of the likelihood function, based on MLE~\cite{jaromir-pra-2001}. 

To define these loss functions in the SGD-QMT framework, we first introduce an abstract representation of the measurement apparatus, modeled by a set of POVM elements ${\Pi} = \{ \Pi_i \}_{i=1}^k$, expressed as a parameterized mapping $\mathcal{P}_{\Pi}(\boldsymbol{\theta})$, such that $\mathcal{P}_{\Pi}(\theta_i) = \Pi_i$ and $\boldsymbol{\theta} = \{ \theta_i \}$. In this representation, the MSE and MLE loss functions can be formally defined as
%
\begin{align}
\mathcal{L}_{\text{MSE}}[\mathcal{P}_{\Pi}(\boldsymbol \theta)] &= \text{Avg}\sum_{i,j} \mleft[ p_{ij} - \text{Tr} \mleft[\mathcal{P}_{\Pi}(\theta_i) \rho_j \mright] \mright]^2 ,
\label{mse-loss}
\\
\mathcal{L}_{\text{MLE}}[\mathcal{P}_{\Pi}(\boldsymbol \theta)] &= \text{Avg}\sum_{i,j} -p_{ij} \log \mleft( \text{Tr} \mleft[\mathcal{P}_{\Pi}(\theta_i) \rho_j \mright] \mright) .
\label{mle-loss}
\end{align}
%
The notation `Avg' in Eqs.~(\ref{mse-loss}) and (\ref{mle-loss}) denotes the average over all POVM-state index pairs $(i,j)$ appearing in the summation. The data $\{ p_{ij}\}$ represents experimentally obtained probabilities, while $\text{Tr} \big[\mathcal{P}_{\Pi}(\theta_i) \rho_j \big]$ denotes the corresponding estimated (predicted) probabilities, with $\{\rho_j \} $ representing the set of input quantum states.



\subsubsection{Gradient-descent algorithms}
\label{sec:GD-Algorithms}

Within the mini-batch SGD framework, we adopt two optimization strategies depending on the parameterization scheme: vanilla gradient descent (VGD) is used for the SM parameterization, while the momentum-based \textit{Adam} optimizer with an adaptive learning is employed for the HONEST parameterization. In both cases, the gradient of the loss function $\mathcal{L}$ with respect to the parameter vector $\boldsymbol{\theta}$ at iteration $t$, denoted by $\mathcal{G}_{\boldsymbol{\theta}^t} = \nabla_{\boldsymbol{\theta}^t} \mathcal{L}$, is computed as
%
\begin{equation}
\mathcal{G}_{\boldsymbol{\theta}^t} = \nabla_{\boldsymbol{\theta}^t} \mathcal{L} \mleft( \mleft[\{\rho_j \}^{(m)}_t, \{\Pi_i \}^{(n)}_t, \{p_{ij}\}_t \mright]; \mathcal{P}_{\Pi}(\boldsymbol \theta^t) \mright) .
\label{eq:gradient}
\end{equation}
The mini-batch $\mleft[\{\rho_j \}^{(m)}_t, \{\Pi_i \}^{(n)}_t, \{p_{ij}\}_t \mright]$ consists of $m$ randomly selected input states $\{\rho_j \}^{(m)}_t$, $n$ randomly selected POVM elements $\{\Pi_i \}^{(n)}_t$, and the corresponding subset of experimental data $\{p_{ij}\}_t$, yielding a total mini-batch size of $m \cdot n$. At each iteration, the gradient update is performed using a freshly sampled mini-batch. This mini-batch stochastic approach efficiently handles large data sets while accelerating convergence by helping the optimizer escape local minima and saddle points, thereby increasing the chances of reaching a global minimum~\cite{dimitri-siam-2000, simon-arxiv-2019}. We refer to \appref{app:hyperparameters} for a more detailed discussion of gradient-descent algorithms and their hyperparameters.

In the following Secs.~\ref{sec:sm} and \ref{sec:honest}, we present valid POVM parameterizations $\mathcal{P}_{\Pi}(\boldsymbol \theta)$ based on the SM and HONEST formalisms, respectively, implemented within the mini-batch SGD framework. We collectively refer to these approaches as SGD-QMT methods.


\subsubsection{SGD-QMT with Stiefel-manifold parameterization}
\label{sec:sm}

The Stiefel manifold (SM) optimization framework has recently emerged as a powerful tool across a variety of quantum technology applications~\cite{Luchnikov2021, Tagare2011, Boumal2023}. In Ref.~\cite{Luchnikov2021}, the authors demonstrate its use in analyzing the low-energy spectrum and eigenstates of multipartite Hamiltonians, performing variational searches for tensor networks in the form of the multiscale entanglement renormalization ansatz, preparing highly entangled arbitrary states, and decomposing quantum gates. Beyond these applications, SM-based optimization has also been successfully employed in quantum tomography tasks, including QST~\cite{gaikwad-arxiv-2025}, QPT~\cite{ahmed-prl-2023}, QMT~\cite{ilia-scipost-2021}, and gate set tomography (GST)~\cite{roth-prxquant-2023}.  In these settings, SM-based parameterizations are used to represent quantum states, processes, measurements, and gate sets in a way that naturally supports gradient-based optimization while preserving essential physical constraints such as orthogonality, complete positivity, and trace preservation~\cite{wen-mp-1013, jiang-mp-2015, absil2009optimization}. Notably, Ref.~\cite{ilia-scipost-2021} introduces a Python library built on TensorFlow that utilizes SM-based gradient optimization, offering a flexible platform for quantum tomography tasks. However, that work provides limited details on the specific parameterizations of the respective entities, and lacks the in-depth analysis required to benchmark these tools when applied to DV and CV systems. 

Here, we extend the use of SM parameterization to efficiently represent sets of POVM elements that satisfy both positivity and completeness constraints during SGD optimization. Within the SM framework, positivity of the POVM elements is ensured via a \textit{Cholesky decomposition}~\cite{white-pra-2001, gaikwad-arxiv-2025}, while the completeness condition is intrinsically enforced by the orthonormality property of the SM.

To formalize this parameterization, consider a set of POVM elements $\{ \Pi_1, \Pi_2, \dots, \Pi_k \}$ characterizing a measurement device in a $d$-dimensional Hilbert space. We first construct a set of $k$ arbitrary complex matrices $ \{T_i\}_{i=1}^k$, with each $T_i \in \mathbb{C}^{d \times d}$. Each POVM element is then expressed as $\Pi_i = T_i^\dagger T_i$, which guarantees that $\Pi_i$ is positive semi-definite. To further enforce the completeness condition $\sum_i \Pi_i = I$, we vertically stack the matrices $T_i$ into a single matrix as
\begin{equation}
\mathbb{T}_k = \begin{bmatrix}
T_1 & T_2 & \cdots & T_k
\end{bmatrix}^{T} ,
\label{eq:sm}
\end{equation}
such that $\mathbb{T}_k \in \mathbb{C}^{k\cdot d \times d}$.
The completeness condition can now be written as
\begin{equation}
\mathbb{T}_k^\dag \mathbb{T}_k =
\begin{bmatrix}
T_1^\dag & T_2^\dag & \cdots & T_k^\dag
\end{bmatrix}
\begin{bmatrix}
T_1 \\[4pt]
T_2 \\[4pt]
\vdots \\[4pt]
T_k
\end{bmatrix}
= I^{d \times d}.
\end{equation}
This orthonormality condition defines the complex Stiefel manifold $St(k \cdot d, d)$ as
\begin{equation}
St(k \cdot d, d) = \mleft\{ \mathbb{T}_k \in \mathbb{C}^{k \cdot d \times d} \mid \mathbb{T}_k^\dag \mathbb{T}_k = I^{d \times d} \mright\} .
\label{eq:sm_def}
\end{equation}

With this, the valid SM parameterization for a given set of POVM elements becomes
\begin{equation}
\{ \Pi_i \}_{i=1}^k = \mathcal{P}_{\Pi}(\mathbb{T}_k) = \mleft\{ \mathbb{T}_k[i]^\dag \mathbb{T}_k[i] \mright\}_{i=1}^{k} ,
\label{eq:sm_para}
\end{equation}
where $\mathbb{T}_k[i] = T_i$ with $\mathbb{T}_k \in St(k \cdot d, d)$ and the loss functions $\mathcal{L}_{\text{MSE}}[\mathcal{P}_{\Pi}(\mathbb{T}_k)]$ and $\mathcal{L}_{\text{MLE}}[\mathcal{P}_{\Pi}(\mathbb{T}_k)]$ are given as
%
\begin{align}
\mathcal{L}_{\text{MSE}} &= \text{Avg}\sum_{i,j} \mleft[ p_{ij} - \text{Tr}\mleft( \mathbb{T}_k[i]^\dag \mathbb{T}_k[i] \rho_j \mright) \mright]^2 , 
\label{mse-sm}
\\
\mathcal{L}_{\text{MLE}} &= \text{Avg}\sum_{i,j} -p_{ij} \log \mleft( \text{Tr} \mleft[\mathbb{T}_k[i]^\dag \mathbb{T}_k[i] \rho_j \mright] \mright) .
\label{mle-sm}
\end{align}
Iteratively optimizing $\mathbb{T}_k$ on the SM that minimizes the given loss function consistently preserves the manifold structure as defined in \eqref{eq:sm_def}, resulting in a valid POVM reconstruction. 

The optimization procedure on the SM can be described briefly in three main steps~\cite{wen-mp-1013, jiang-mp-2015, absil2009optimization}:
\begin{enumerate}
\item At each iteration, we compute and reconstruct the three quantities
\begin{align}
\tilde{\mathcal{G}}_{\mathbb{T}_k} &= \mathcal{G}_{\mathbb{T}_k}/{\mleft\| \mathcal{G}_{\mathbb{T}_k}\mright\|}_{l_2},
\\
\mathbb{A} &= \begin{bmatrix} \Tilde{\mathcal{G}}_{\mathbb{T}_k} & \mathbb{T}_k  \end{bmatrix} ,
\\
\mathbb{B} & = \begin{bmatrix} \mathbb{T}_k & -\tilde{\mathcal{G}}_{\mathbb{T}_k}   \end{bmatrix} ,
\end{align}
where $\mathcal{G}_{\mathbb{T}_k} = \nabla_{\mathbb{T}_k} \mathcal{L}$, as defined in \eqref{eq:gradient}, denotes the mini-batch gradient of the loss function with respect to $\mathbb{T}_k$.
\item Using the Cayley transform and the Sherman--Morrison--Woodbury formula, we compute the conjugate gradient $\mathcal{G}^*_{\mathbb{T}_k}$ as~\cite{jun-arxiv-2020}
\begin{equation}
\mathcal{G}^*_{\mathbb{T}_k} = \mathbb{A} \left( I + \frac{\eta}{2} \mathbb{B}^{\dagger} \mathbb{A} \right)^{-1} \mathbb{B}^{\dagger} \mathbb{T}_k ,
\label{eq:retraction}
\end{equation}
where the scalar $\eta$ is the step size (learning rate).
\item Finally, applying the VGD update rule, the next iterative update on the SM is computed as~\cite{ahmed-prl-2023}
\begin{equation}
\mathbb{T}_k^{t+1} = \mathbb{T}_k^{t} - \eta \, \mathcal{G}_{\mathbb{T}_k^t}^{*} .
\label{eq:sm_update}
\end{equation}
\end{enumerate}
This sequence of steps constitutes the \textit{retraction procedure} for performing gradient-based optimization on the SM, which ensures that $\mathbb{T}_k^{t+1} \in St(k \cdot d, d) \: \forall t $. Thus, this procedure results in reconstruction of a valid set of POVM elements, which can be obtained using \eqref{eq:sm_para}.


\subsubsection{SGD-QMT with HONEST Parameterization}
\label{sec:honest}

In this section, we introduce an alternative parameterization of POVM elements based on Hermitian operator normalization via eigenvalue-scaling technique (HONEST). This approach ensures both the positivity and completeness properties required for POVMs, and integrates well with SGD optimization frameworks. The HONEST parameterization involves two key steps: (i) constructing positive semidefinite Hermitian operators using Cholesky decomposition, and (ii) enforcing the completeness condition via a projective normalization step using an eigenvalue-scaling technique.

To properly formalize this parameterization, we begin with a set of arbitrary complex matrices $\{ T_i \}_{i=1}^k$, where each $T_i \in \mathbb{C}^{d \times d}$. Using Cholesky decomposition, we define the unnormalized POVM elements as $T_i^\dag T_i$, ensuring they are positive semidefinite. To further impose the completeness condition, we first compute the sum $S = \sum_i T_i^{\dagger} T_i$ and obtain its eigenvalue decomposition $S = V \Lambda V^{\dagger}$, where $V$ contains the orthonormal eigenvectors of $S$, and $\Lambda = \mathrm{diag}(\lambda_1, \lambda_2, \dots, \lambda_d)$ is a diagonal matrix of its real and non-negative eigenvalues. The inverse square root of $S$ is then computed via eigenvalue scaling as
\begin{equation}
S^{-1/2} = V \Lambda^{-1/2} V^\dag \quad \text{with} \quad \lambda_i \rightarrow \frac{1}{\sqrt{\lambda_i}} .
\label{eq:eig}
\end{equation}

Note that, in practice, if an eigenvalue $\lambda_i \rightarrow 0$, its inverse square root diverges, i.e., $\frac{1}{\sqrt{\lambda_i}} \rightarrow \infty$. Furthermore, although the matrix $S$ is theoretically positive semidefinite, numerical inaccuracies arising from finite-precision arithmetic may introduce small negative eigenvalues. This can result in undefined or complex values during the computation of $\frac{1}{\sqrt{\lambda_i}}$. To address these issues and ensure numerical stability, we regularize the spectrum of $S$ by clipping its eigenvalues from below using a small positive threshold, typically $\delta = 10^{-8}$, prior to performing the inverse square root operation.

Using \eqref{eq:eig}, we sandwich the unnormalized POVM elements as
\begin{equation}
{\Pi}_i = {S^{-1/2}}^\dag T_i^\dag T_i S^{-1/2} .
\label{eq:sandwich}
\end{equation}
The positive semidefinite constraint of each $ \Pi_i$ is enforced by \eqref{eq:sandwich} and easily verified as
\begin{equation}
\langle \phi \vert {S^{-1/2}}^\dag T_i^\dag T_i S^{-1/2} \vert \phi \rangle \geq 0 \quad \forall \ket{\phi} ,
\end{equation}
and the completeness condition can be proven as
\begin{align}
\sum_i {\Pi}_i & = {S^{-1/2}}^\dag \mleft( \sum_i  T_i^\dag T_i \mright) S^{-1/2} 
\nonumber
\\
& = S^{-1/2} S S^{-1/2} = I.
\label{eq:honest_completeness}
\end{align}
Note that, since $S$ is a Hermitian positive semidefinite matrix, we have $S^{-1/2} = {S^{-1/2}}^\dag$, which is used in \eqref{eq:honest_completeness}.

Thus, the final HONEST Parameterization for measurement devices characterized by a set of POVM elements can be given as
\begin{equation}
\{ {\Pi}_i \}_{i=1}^k = \mathcal{P}_{\Pi} \mleft(\{ T_i \}_{i=1}^k \mright) = \mleft\{ S^{-1/2} T_i^\dag T_i S^{-1/2} \mright\}_{i=1}^{k}
\end{equation}
and the loss functions $\mathcal{L}_{\text{MSE}}\mleft[\mathcal{P}_{\Pi}\mleft(\{ T_i \}_{i=1}^k \mright) \mright] $ and $\mathcal{L}_{\text{MLE}}\mleft[\mathcal{P}_{\Pi}\mleft(\{ T_i \}_{i=1}^k \mright) \mright] $ become
\begin{align}
\mathcal{L}_{\text{MSE}} &= \text{Avg} \sum_{i,j} \mleft[ p_{ij} - \text{Tr} \mleft[S^{-1/2} T_i^\dag T_i S^{-1/2} \rho_j \mright] \mright]^2 , 
\label{mse-honest}
\\
\mathcal{L}_{\text{MLE}} &= \text{Avg} \sum_{i,j} -p_{ij} \log \mleft( \text{Tr} \mleft[S^{-1/2} T_i^\dag T_i S^{-1/2} \rho_j \mright] \mright) .
\label{mle-honest}
\end{align}
Here, we employ the momentum-based Adam optimizer to iteratively minimize the given loss function with respect to $\{ T_i \}_{i=1}^k$. The gradient of the loss function required by the Adam algorithm is computed according to \eqref{eq:gradient}. 

The complete SGD-QMT parameter-update procedure using the HONEST parameterization can thus be summarized as
\begin{equation}
\{ T_i^t \}_{i=1}^k \xrightarrow{\text {Adam}} \{ T_i^{t+1} \}_{i=1}^k \xrightarrow{\text {HONEST}} \{ \tilde{T}_i^{t+1} \}_{i=1}^{k} .
\end{equation}
Here, the HONEST step is implemented using Eqs.~(\ref{eq:eig}) and (\ref{eq:sandwich}) to obtain
\begin{equation}
\tilde{T}_i^{t+1} = T_i^{t+1} \mleft(S^{-1/2}\mright)^{t+1} .
\label{eq:honest-step}
\end{equation}
Note that $\mleft(S^{-1/2}\mright)^{t+1}$ in \eqref{eq:honest-step} is computed according to \eqref{eq:eig} using Adam-updated matrices $\{T_i^{t+1}\}_{i=1}^k$. The resulting matrices $\{\tilde{T}_i^{t+1}\}_{i=1}^k$ are then used to evaluate the corresponding valid POVM elements as
\begin{equation}
{\Pi}_i^{t+1} = \mleft( \tilde{T}_i^{t+1} \mright)^\dag \tilde{T}_i^{t+1} ,
\end{equation}
satisfying ${\Pi}_i^{t+1} \geq 0 \: \forall i,t$ and $\sum_i {\Pi}_i^{t+1} = I \: \forall t$.

\subsection{Convex optimization methods}
\label{sec:cco}

Constrained convex optimization (CCO) methods have emerged as the standard and widely adopted framework for QMT~\cite{Schapeler-qst-2025}, formulated typically in the form of a least-squares optimization subject to physical constraints such as positivity and completeness of the POVM elements. The least-squares CCO problem for QMT can be simply described as
\begin{subequations}
\begin{align} 
\min_{ \{\Pi_i^{\text{est}} \} } \sum_{i,j} \mleft[ p_{ij} - \text{Tr}( \Pi_i^{\text{est}} \rho_j ) \mright]^2 
\label{eq:ls-cco} 
\\
\text{s.t.} \quad \Pi_i^{\text{est}} \geq 0 \quad \forall i ,
\\ 
 \sum_i \Pi_i^{\text{est}} = I  ,
\end{align}
\end{subequations}
%
where $p_{ij}$ denotes the experimentally obtained probability, and $\text{Tr}(\Pi_i^{\text{est}} \rho_j)$ represents the corresponding estimated probability. 


The least-squares CCO problem, as presented in \eqref{eq:ls-cco}, can be straightforwardly compiled within well-established convex optimization tools such as CVX~\cite{diamond2016cvxpy} or YALMIP~\cite{lofberg-2004}, and subsequently solved efficiently using their integrated solvers~\cite{cvx-solvers}. However, these tools offer limited flexibility in terms of data processing and become inadequate for large-scale systems, often demanding extensive computational time.


\subsection{Additional customizations}
\label{sec:customization}

The performance and efficiency of QMT algorithms can be further improved by incorporating various regularization techniques, as discussed in Ref.~\cite{xiao-automatica-2023}. Among these, three key customizations can be introduced into the main algorithms described here in Sec.~\ref{sec:Methods}, depending on the available prior knowledge or the specific tomography objective. These modifications may lead to substantial improvements, including reduced computational overhead and data requirements with improved estimation accuracy:
\begin{itemize}
\item \textit{Rank-controlled ansatz:}
The optimization complexity in SGD-QMT algorithms can be significantly reduced by adopting a rank-controlled ansatz for each POVM element, parameterized using $\{ T_i \in \mathbb{C}^{r_i \times d} \}_{i=1}^k$, where $r_i$ is the rank of $i$th POVM element. For example, in the case of rank-1 Pauli projective measurements, one can employ $\{ T_i \in \mathbb{C}^{1 \times d} \}_{i=1}^k$. This reduces the number of parameters to be optimized from $k \cdot d^2$ (full rank) to $k \cdot d$ (rank 1), thereby accelerating the optimization process. Note that while all rank-1 POVM operators represent projective measurements, not all projective measurements correspond to rank-1 POVMs, e.g., photon detection.
\item \textit{L1-norm regularization:}
An additional L1 term, $\sum_i\vert \vert \mathcal{P}_{\Pi}( \theta_i) \vert\vert_{l_1}$, can be incorporated into the base loss function (MSE or MLE) to promote sparsity in the reconstructed POVMs. This regularization is particularly advantageous in scenarios with highly limited datasets, where the underlying solution is expected to be sparse, as demonstrated in compressed-sensing methods~\cite{rod-prb-2014, gaikwad-quntinfo-2022}. By driving small-valued parameters towards zero, the L1 penalty emphasizes the most significant components of the solution, improving robustness and interpretability.
\item \textit{Nuclear norm regularization:}
Including a nuclear norm penalty, $\sum_i\vert \vert \mathcal{P}_{\Pi}( \theta_i) \vert\vert_{*}$, in the base loss function promotes low-rank structure in the reconstructed POVMs. This regularization biases the optimization towards solutions with fewer significant singular values. Such a constraint is particularly beneficial when the underlying measurement operators are expected to be intrinsically low-rank (e.g., projective or near-projective measurements), thereby improving the physical plausibility and efficiency of the reconstruction using relatively reduced data sets.
\end{itemize}
%


\subsection{Data sets and performance metrics}
\label{sec:datasets}

In this section, we briefly describe various measurement processes (which act as data sets) and performance metrics that we use to benchmark our SGD-QMT algorithms for DV and CV systems.


\subsubsection{Discrete-variable systems}

For $N$-qubit DV systems, we examine two types of measurement processes: (i) randomly generated POVM elements (all assumed to be full rank), with varying number and size, and (ii) Pauli projective measurements, represented by a complete set of $2^N$ rank-1 POVM elements, constructed from the orthonormal eigenvectors $\{ \ket{e_i} \}$ of given Pauli operators as $\{\Pi_i = \lvert e_i\rangle\langle e_i\rvert\}$ with an in-depth analysis of projective measurements in the computational basis (i.e., the $\sigma_z^{\otimes N}$ basis). Furthermore, the probe input states used during the data acquisition phase are sampled from the following set:
\begin{equation}
\rho = \left\{ 
\begin{bmatrix} 1 & 0 \\ 0 & 0 \end{bmatrix},\ 
\begin{bmatrix} 0 & 0 \\ 0 & 1 \end{bmatrix},\ 
\frac{1}{2}\begin{bmatrix} 1 & 1 \\ 1 & 1 \end{bmatrix},\ 
\frac{1}{2}\begin{bmatrix} 1 & -i \\ i & 1 \end{bmatrix} 
\right\}^{\otimes N} .
\label{eq:input_states}
\end{equation}
%


\subsubsection{Continuous-variable systems}

For CV systems, we focus on two physically significant and relevant measurement scenarios: (i) photon detection, modeled by two POVM elements, and (ii) photon counting, characterized by a POVM set whose cardinality scales linearly with the dimension of the truncated Hilbert space. For the data-acquisition step, the probe input states are selected from a set of coherent states $\{ \ket{\alpha} \}$, defined as
\begin{equation}
\ket{\alpha} = e^{-\frac{1}{2}|\alpha|^2} \sum_{n=0}^{\infty} \frac{\alpha^n}{\sqrt{n!}} \ket{n} ,
\label{eq:coherent_state}
\end{equation}
where $\alpha = r e^{i \phi}$ denotes a complex amplitude. It is important to note that the number of coherent states required for QMT usually scales quadratically with the dimension of the truncated Hilbert space.  


\subsubsection{Performance metrics}

To assess the overall performance of our QMT algorithms, we employ two different benchmarking metrics that quantify the difference between two sets of POVM elements---typically the reference (target) POVMs $\{ \Pi_i^{\text{ref}}\}$ and those estimated (predicted) $\{ \Pi_i^{\text{est}}\}$ by the algorithm under evaluation. These metrics are: (i) the average Frobenius norm $\bar{\mathcal{F}} = \text{Avg}(\mathcal{F})$ over POVM elements~\cite{julia-prr-2025}, and (ii) the average Wasserstein distance $\bar{\mathcal{W}} = \text{Avg}(\mathcal{W})$ over probe input states~\cite{yoav-anrev-2019}.  

The Frobenius norm $\mathcal{F}$ between two operators $A \in \{ \Pi_i^{\text{ref}}\}$ and $B \in \{ \Pi_i^{\text{est}}\}$ is defined as follows:
\begin{equation}
{\mathcal{F}}( A,  B ) = \text{Tr}\mleft[ (A - B)^{\dagger}(A - B) \mright] , 
\label{frob}
\end{equation}
%
Unlike the Frobenius norm, which measures the difference between two operators, the Wasserstein distance $\mathcal{W}$ quantifies the difference between two probability distributions. Note that, in the QMT context, the dataset $\{ p_{ij} \}_{i=1}^k \: \forall j $ represents valid probability distributions.

For a given probe input state $\rho_j$, we compute the Wasserstein distance $\mathcal{W}$ between the reference cumulative probability distribution $\mathbf{p}_j^{\rm ref}$, derived from the true POVM elements $\{ \Pi_i^{\text{ref}} \}$, and the estimated cumulative probability distribution $\mathbf{p}_j^{\rm est}$, obtained from the estimated POVM elements $\{ \Pi_i^{\text{est}} \}$:
\begin{equation}
\mathcal{W}(\textbf{p}_j^{\rm ref}, \textbf{p}_j^{\rm est} ) = \sum_{i=1}^{k-1} \vert \textbf{p}_j^{\rm ref} (x_i) - \textbf{p}_j^{\rm est} (x_i) \vert \cdot (x_{i+1}-x_i) ,
\end{equation}
where $x_i$'s are the position (POVM) labels: $x_i = i$, where these probabilities are assigned.

It is worth noting that one may also adopt the conventional readout fidelity metric---particularly relevant for measurements in the computational basis---which is commonly defined as the detector's average success probability for correctly identifying computational basis states $\ket{e_i}$, with $\ket{e_i} \in \{ \ket{0}, \ket{1} \}^{\otimes N}$. This metric is widely used for calibrating measurement hardware. However, we emphasize that readout fidelity only captures information about the diagonal entries of the reconstructed POVM elements and therefore offers a limited view of the measurement apparatus model. For this reason, we adopt the Frobenius (and Wasserstein) norm as our benchmarking metric, as it provides a global measure of discrepancy by accounting for contributions from all elements of the POVM matrices.


\section{Results}
\label{sec:Results}

In this section, we compare our various SGD-QMT methods with CCO methods using CVX, as outlined in Sec.~\ref{sec:Methods}. Here, the SGD-QMT category includes SM-MSE (Stiefel manifold parameterization with MSE loss), SM-MLE (Stiefel manifold parameterization with MLE loss), HONEST-MSE (HONEST parameterization with MSE loss), and HONEST-MLE (HONEST parameterization with MLE loss).

We implement these QMT algorithms numerically on various data sets described in Sec.~\ref{sec:datasets} and evaluate their performance. Our analysis emphasizes computational time complexity, reconstruction fidelity (based on the metrics described in Sec.~\ref{sec:datasets}), and the effects of noisy probe states (shown in \appref{app:noisy data}) in both DV and CV systems. 


\subsection{Discrete-variable systems}
\label{sec:dv_results}

We begin by analyzing the performance of the QMT methods described in Sec.~\ref{sec:Methods} for DV systems comprising up to six qubits (limited by memory constraints). For the DV case, we perform numerical simulations in two representative scenarios: (i) randomly generated full-rank POVM elements (\secref{sec:Results-DV-FullRank}), and (ii) rank-1 Pauli projective measurements (\secref{sec:Results-DV-Rank1}). 

In addition to these general benchmarks, we conduct a focused case study of SGD-QMT methods on projective measurements in the computational basis,
$\Pi = \{ \ket{0}\bra{0}, \ket{1}\bra{1} \}^{\otimes N}$ (\secref{sec:Results-DV-Projective}).
This setting is of particular practical relevance, as most existing quantum computing hardware platforms employ fixed measurement setups that are restricted to the computational basis. By analyzing this scenario in depth, we aim to evaluate how well SGD-QMT protocols perform under experimentally realistic conditions.




\subsubsection{Full-rank random POVM sets}
\label{sec:Results-DV-FullRank}

\begin{figure*}[t]
\centering
\includegraphics[width=\linewidth]{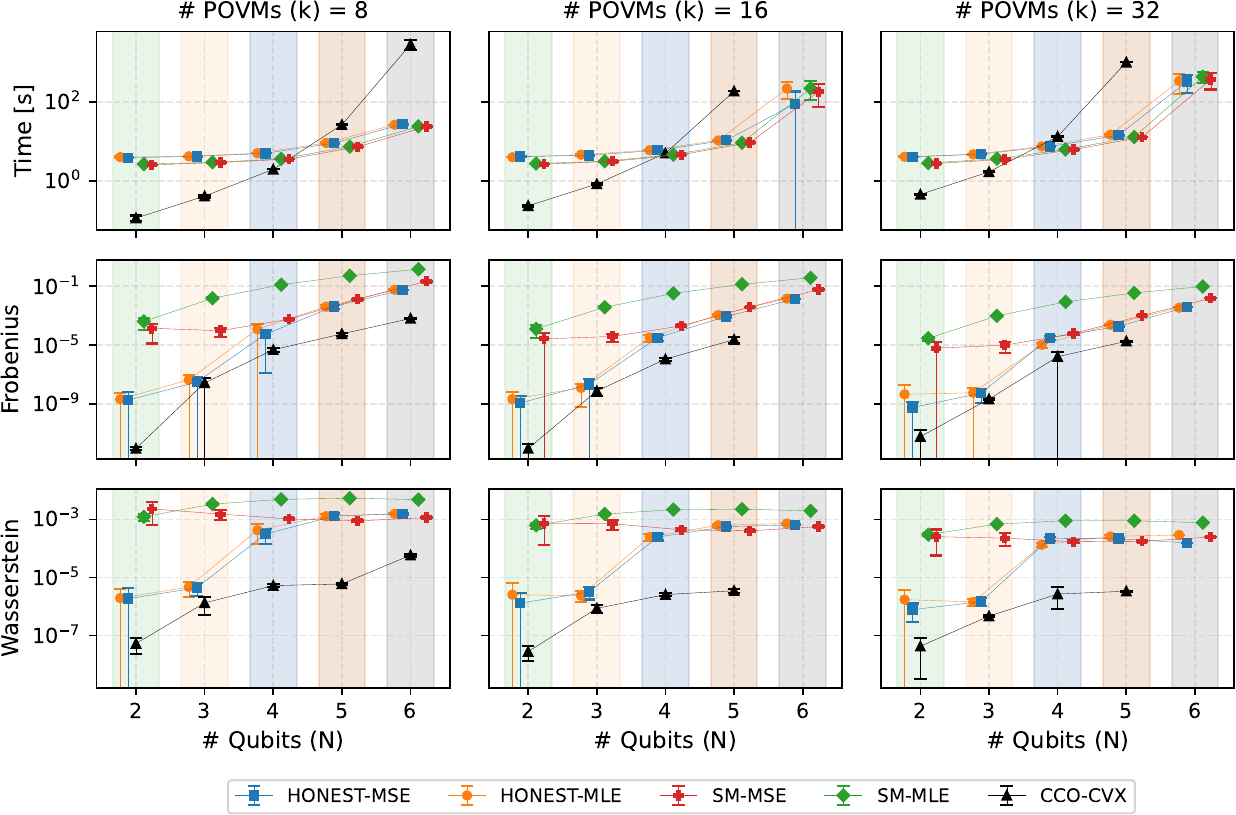}
\caption{Time complexity and performance metrics of different QMT data-processing methods for random full-rank POVM sets. The upper and lower lines for each marker are error bars denoting one standard deviation. The first row shows runtime in seconds, while the second and third rows report the Frobenius distance and the Wasserstein distance, respectively, for the reconstructed POVMs. Each metric on the $y$ axes is averaged 15 randomly generated POVM sets of the indicated size. The first column corresponds to QMT with $k=8$ POVM elements, while second and third columns correspond to $k=16$ and $k=32$, respectively. In all cases, the $x$ axis denotes the number of qubits, with the maximum number of iterations fixed at 2000. The data points for given qubit is confined within corresponding faint colored vertical band to avoid value overlap. }
\label{fig:time_complexity}
\end{figure*}

Figure~\ref{fig:time_complexity} summarizes the performance of the SGD-QMT methods HONEST-MSE (teal), HONEST-MLE (orange), SM-MSE (red), and SM-MLE (green), and compares them against the state-of-the-art CCO-CVX approach (black). In all figure panels, the $x$-axis represents the number of qubits, with each value highlighted by a faint colored vertical band. All data points corresponding to a given qubit number are confined within the associated vertical band, providing clear visual separation and preventing overlap between values. The first row of \figref{fig:time_complexity} shows the total computational time (in seconds) required to finish 2000 iterations and reconstruct the full POVM set. The second and third rows report the corresponding Frobenius norm and Wasserstein distance, respectively, which serve as benchmarking metrics for assessing the reconstruction quality. All three metrics on the $y$ axes are averaged over $15$ randomly generated POVM sets of the specified sizes for each column. 

The numerical results in \figref{fig:time_complexity} indicate that, for smaller systems of up to four qubits (with the number of POVM elements as small as $k=8$), the CCO-CVX method consistently outperforms all SGD-QMT variants, offering both faster computational times and superior reconstruction quality. However, this advantage does not scale with system size. As the number of qubits ($N$) and POVM elements ($k$) increases, the computational cost of CCO-CVX grows rapidly, making it impractical for larger systems (high qubit number or many POVM elements). This trend is clearly visible in the first row of \figref{fig:time_complexity}, where the runtime of CCO-CVX increases steeply with $N$ and $k$. In contrast, SGD-QMT methods scale much more favorably: for four or more qubits with eight or more POVM elements, they achieve speedups of up to a few orders of magnitude compared to CCO-CVX. For instance, in the case of $N=5$ and $k=32$, CVX requires roughly 1000 seconds ($\approx 15$ minutes) to reconstruct the complete POVM set, achieving an average Frobenius error of $\sim 10^{-5}$. In contrast, SGD-QMT methods reconstruct the full POVM set in just a few tens of seconds---almost 100 times faster---while attaining a comparable average Frobenius error around $\sim 10^{-4}$ when using the HONEST parameterization. Moreover, for six-qubit systems with $k=16$ or higher, the CCO-CVX approach fails to obtain a solution within a reasonable time frame, underscoring its limitations for larger problem instances, whereas all four SGD-QMT algorithms return solutions in approximately two minutes for $N=6$ and $k=32$. 

However, this dramatic gain in computational efficiency does come at the expense of slightly worse reconstruction quality, as evidenced by the Frobenius and Wasserstein metrics (the second and third rows in \figref{fig:time_complexity}), where SGD-QMT have slightly or clearly lower accuracy, in some cases approaching that of CCO-CVX. The gap between SGD-QMT and CCO-CVX is larger in terms of the average Wasserstein distance, but that distance still consistently falls within the range of $10^{-5}-10^{-3}$ or better for the SGD-QMT methods, ensuring sufficiently high accuracy for most practical purposes. Note that the maximum number of iterations for all SGD-QMT methods in all cases presented in \figref{fig:time_complexity} is set to 2000 for consistent comparison. The estimation accuracy can be further improved by increasing the iteration count for high-dimensional systems. These results highlight the scalability advantage of SGD-QMT methods, particularly in regimes where traditional convex optimization becomes computationally prohibitive.

While the runtime of all SGD-QMT algorithms are nearly identical, it is clear from \figref{fig:time_complexity} that, in almost all scenarios, the HONEST parameterization yields substantially better reconstruction quality, with Frobenius norm and Wasserstein distance values smaller than those obtained with the SM parameterization. Furthermore, for the SM case, the MSE loss function outperforms the MLE loss function, whereas under the HONEST parameterization, the MSE and MLE loss functions exhibit nearly identical performance. 


The results shown in \figref{fig:time_complexity} further reveal that the time complexity of the SGD-QMT algorithms is driven primarily by the number of qubits and not so much by the number of POVM elements; a clear difference compared to the CCO-CVX method, whose runtime is strongly affected by both. Specifically, for a fixed $N$, the runtimes for the SGD-QMT algorithms remain comparable across all tested values of $k = 8$, 16, and 32, while it increases from about $10^{1}$ seconds to $10^{3}$ seconds when $k$ is increased from 8 to 32 for $N=5$ with CCO-CVX. In contrast, when the number of qubits increases for a fixed $k$, the computational cost grows substantially for both SGD-QMT and CCO-CVX algorithms. This scaling behavior suggests that SGD-QMT algorithms are robust against variations in measurement settings (varying numbers of POVM elements), but face their main computational bottleneck in system size, making them especially suitable for scenarios with large numbers of POVM elements but moderate qubit counts. At the same time, the steep scaling for the CCO-CVX method limits its practicality even for relatively small qubit systems when a large number of POVM elements are considered.

It is worth noting that the performance of the SGD-QMT methods likely can be further enhanced by optimizing various hyperparameters (see also \appref{app:hyperparameters}) like batch size, learning rate, and increasing iteration count, in addition to the customizations discussed in Sec.~\ref{sec:customization}.


\subsubsection{Rank-1 Pauli projective measurements}
\label{sec:Results-DV-Rank1}

\begin{figure}
\centering
\includegraphics[width=0.95\linewidth]{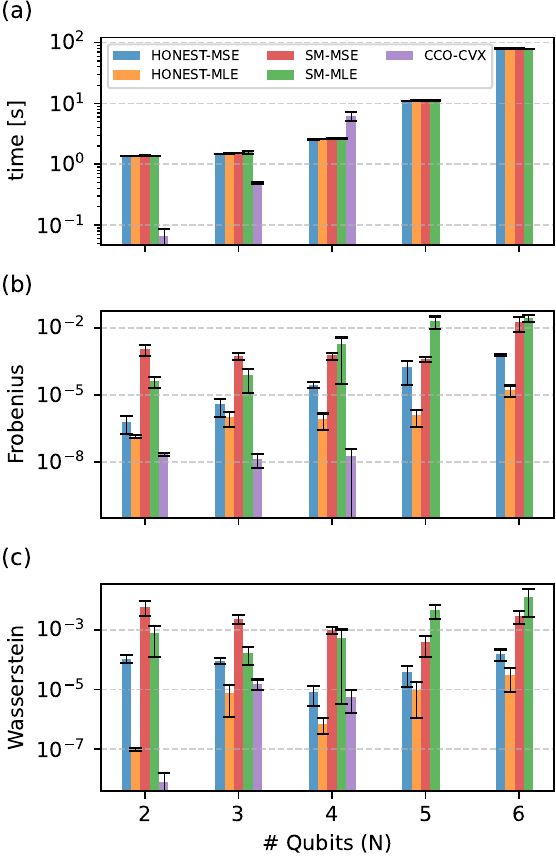}
\caption{SGD-QMT performance for rank-1 Pauli projective measurements. 
(a) Runtime (seconds),
(b) Frobenius distance, and
(c) Wasserstein distance, all as functions of qubit number $N$.
The bar plots show HONEST-MSE (teal), HONEST-MLE (orange), SM-MSE (red), SM-MLE (green) and CCO-CVX (purple, up to four qubits). All three performance metrics are averaged over 15 randomly generated Pauli projective measurements. The error bars represent standard deviation across these instances. The maximum number of iterations is fixed at 1000 in all cases.}
\label{fig:time_complexity_pauli}
\end{figure}
\begin{figure*}[t]
\centering
\includegraphics[width=\linewidth]{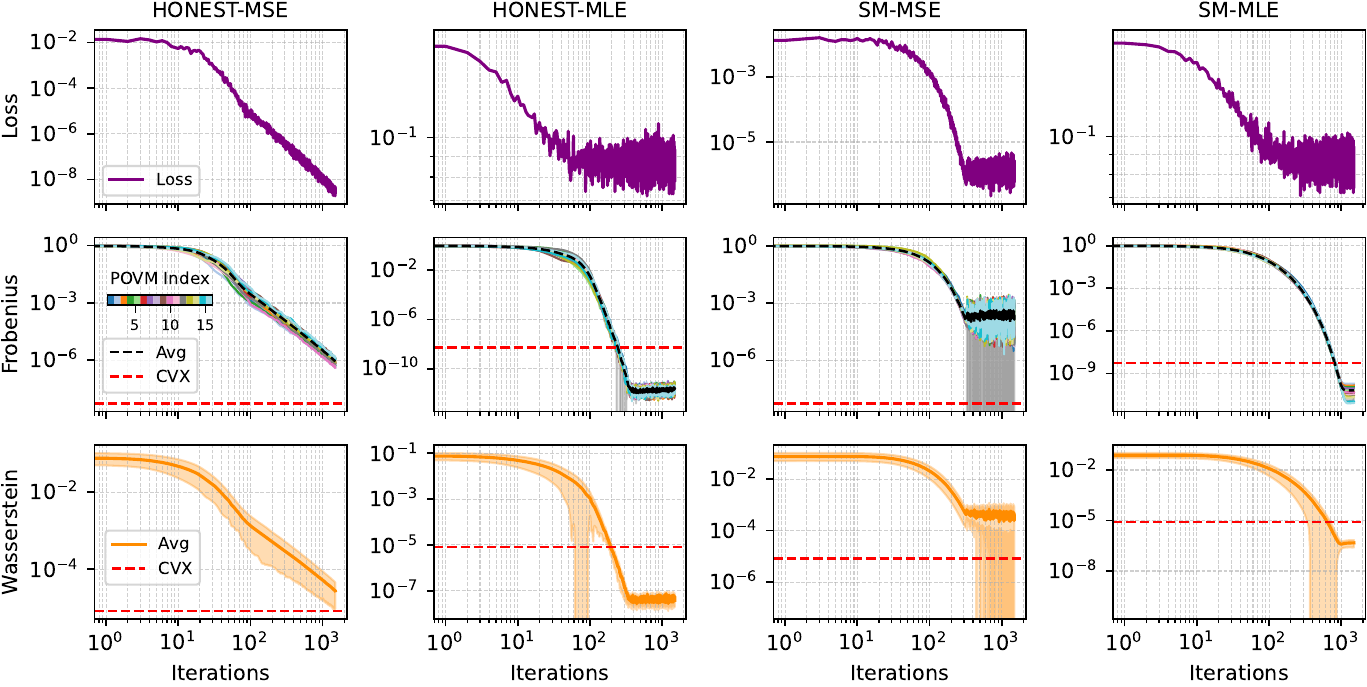}
\caption{Performance of SGD-QMT data-processing algorithms for four-qubit projective measurements in the computational basis $\Pi = \{ \ket{0}\bra{0}, \ket{1}\bra{1} \}^{\otimes 4}$, evaluated with respect to loss (first row), Frobenius norm (second row) and Wasserstein distance (third row), as a function of the number of iterations. The columns correspond to results for HONEST-MSE, HONEST-MLE, SM-MSE, and SM-MLE, in that order. In the second row, each POVM element's Frobenius distance is plotted in a distinct color, with the black dashed line indicating the average Frobenius norm over all 16 POVM elements and gray shading indicating one standard deviation for that average. In the third row, the solid orange line denotes the average Wasserstein distance computed over 16 input states, and the shaded region indicates the standard deviation. In the second and third rows, the results obtained with CCO-CVX are shown as dashed red lines. The maximum number of iterations is set to 1500, with the average total runtime for all SGD-QMT methods below 4 seconds. Note that the $y$ axis scales differ across both rows and columns.}
\label{fig:iteration}
\end{figure*}
Next, we evaluate the performance of our SGD-QMT algorithms for Pauli projective measurements, which are described by $2^N$ rank-1 POVM elements (projectors), computed as an outer product of normalized eigenvectors of given $N$-qubit Pauli matrices. The results shown in \figref{fig:time_complexity_pauli} summarize the comparison of the SGD-QMT methods---HONEST-MSE (cyan), HONEST-MLE (orange), SM-MSE (red), and SM-MLE (green)---with the CCO-CVX method (purple, up to four qubits) in terms of three performance metrics, as a function of the number of qubits: average computational time [\figpanel{fig:time_complexity_pauli}{a}], average Frobenius norm error [\figpanel{fig:time_complexity_pauli}{b}], and average Wasserstein distance [\figpanel{fig:time_complexity_pauli}{c}], each obtained by repeating the reconstruction over multiple randomly chosen Pauli projective measurements. For consistency, the maximum number of iterations in all cases is fixed at 1000 and can be increased further for further improvement. 

The results in \figpanel{fig:time_complexity_pauli}{a} indicate that the computational time complexity across all SGD-QMT algorithms remains nearly identical (as also observed in the case of full-rank random POVM sets in Fig.~\ref{fig:time_complexity}). However, the corresponding reconstruction quality, as quantified by the Frobenius norm and Wasserstein distance in Figs.~\figpanelNoPrefix{fig:time_complexity_pauli}{b} and \figpanelNoPrefix{fig:time_complexity_pauli}{c}, exhibits substantial variation, with differences spanning several orders of magnitude. For two and three qubits, CCO-CVX outperforms SGD-QMT in terms of runtime, requiring an order of magnitude less time while also achieving higher reconstruction quality. However, the efficiency of CCO-CVX deteriorates rapidly beyond three qubits, where it requires orders of magnitude more time than SGD-QMT methods while providing comparable reconstruction accuracy with HONEST-MLE. We report CCO-CVX results only up to four qubits, as for the five-qubit system CCO-CVX failed to produce a solution even after one hour. In contrast, for random POVM sets, CCO-CVX was able to obtain a solution in roughly 15 minutes, as shown in Fig.~\ref{fig:time_complexity}. This significantly limits the practical usability of CVX for systems larger than four qubits. Among the tested SGD-QMT methods, the HONEST-MLE algorithm consistently achieves the highest reconstruction accuracy across all qubit numbers, delivering superior performance without incurring additional computational cost or requiring more iterations. This is different than the results for full-rank random POVM sets in Fig.~\ref{fig:time_complexity}, where HONEST-MSE performed best in terms of reconstruction accuracy, although HONEST-MLE generally was almost just as good. 

Notably, Pauli projective measurements are characterized by rank-1 projectors, making them the maximally sparse class of POVM elements in the respective Pauli eigenbasis (assumed to be known). This structural property can, in principle, be exploited to both reduce computational overhead and enhance reconstruction fidelity by including L1 norm into the loss function and explicitly constraining the reconstructed POVMs to be rank-1 with rank-1 ansatz, as outlined in Sec.~\ref{sec:customization}. Incorporating such prior information effectively tailors the tomography procedure to the underlying POVM structure, leading to more efficient and accurate estimation using heavily reduced data sets. Nevertheless, in the present numerical simulations, we deliberately refrain from imposing this assumption in order to maintain methodological generality. This ensures that the reported performance comparisons reflect the intrinsic capabilities of the algorithms themselves, without relying on problem-specific prior knowledge.

\subsubsection{Projective measurements in the computational basis}
\label{sec:Results-DV-Projective}

In this section, we present an in-depth case study on the performance of our SGD-QMT algorithms for one of the most practically relevant measurement settings: projective measurements in the computational basis (the Pauli-$\sigma_z^{\otimes N}$ basis),
$\Pi = \{ \ket{0}\bra{0}, \ket{1}\bra{1} \}^{\otimes N}$, with $N=4$ taken here as a representative example. In \figref{fig:iteration}, we show how the different SGD-QMT algorithms converge with respect to the number of iterations for this example. We also include the Frobenius norm and Wasserstein distance obtained using the CCO-CVX method (dashed red lines) for reference.

Figure~\ref{fig:iteration} is organized as a $3 \times 4$ grid: the columns correspond to the specific SGD-QMT algorithms and the rows represent the benchmarking metrics (averaged over multiple runs) as functions of the iteration count. To capture the overall convergence behavior, we fix the maximum number of iterations at 1500 for all methods. Notably, the total computational time required to complete 1500 iterations in each case is under 4 seconds, corresponding to a rate of approximately 400 iterations per second or better. This highlights that while the algorithms differ in reconstruction quality, they are all computationally efficient in practice.

The numerical results in \figref{fig:iteration} clearly demonstrate that HONEST-MLE (second column) outperforms all other SGD-QMT methods, achieving average Frobenius norm and Wasserstein distance values of approximately $10^{-11}$ and $10^{-7}$, respectively---several orders of magnitude more accurate than the other three approaches. This performance gain is also particularly pronounced in the case of rank-1 Pauli projective measurements, as further confirmed by the results in \figref{fig:time_complexity_pauli}. In contrast, when applied to fully random full-rank POVM sets, HONEST-MSE and HONEST-MLE exhibit nearly equal performance (\figref{fig:time_complexity}), indicating that the advantage of HONEST-MLE is most significant when the underlying measurement operators possess the rank-1 structure.  Additionally, in all the methods (with the exception of SM-MSE), the error bars--shown as lightly shaded regions around the average Frobenius norm (black dashed line) and average Wasserstein distance (dark orange line)--remain narrow, indicating that each reconstructed POVM element achieves accuracy of the same order as the reported averages. 

We observe a particularly striking feature for HONEST-MLE in \figref{fig:iteration}: the average Frobenius error remains nearly constant at $\sim 10^{-2}$ for the first 100 iterations, but then undergoes a sharp drop, decreasing from $\sim 10^{-2}$ at the 100th iteration to $\sim 10^{-10}$ by around the 300th iteration---an improvement of nearly eight orders of magnitude within just 200 iterations. A qualitatively similar but less pronounced convergence trend is also seen for SM-MLE. In contrast, HONEST-MSE exhibits much slower error reduction (slow convergence): the Frobenius norm decreases only from $\sim 10^{-3}$ at around the 100th iteration to $\sim 10^{-4}$ at around the 300th iteration, corresponding to merely a single order of magnitude improvement over the same interval. This behavior suggests that HONEST-MSE would require significantly more iterations to reach high-accuracy reconstructions comparable to HONEST-MLE. Finally, we note that SM-MSE performs the worst overall, with the average Frobenius error saturating at $\sim 10^{-4}$ and showing no meaningful improvement with increasing iteration count. 


\subsection{Continuous-variable systems}
\label{sec:cv_results}

We now turn to the performance of our SGD-QMT algorithms in the context of CV systems. Our analysis focuses on two of the most practically relevant measurement processes in CV systems: (i) photon detection and (ii) photon counting, which we examine in detail in Secs.~\ref{sec:photon_detection} and \ref{sec:photon_counting}, respectively. In both scenarios, we employ single-mode optical coherent states, as defined in \eqref{eq:coherent_state}, as the probe states. For demonstration purposes, the Hilbert space is truncated (Fock space cutoff) to dimension 32, and the probe states are sampled over a grid of coherent amplitudes $\alpha_m = x_m + i y_m$, where $x_m, y_m \in [-\alpha ,\alpha]$ with 32 equally spaced points along each axis, yielding a total of 1024 distinct coherent probe states.


\subsubsection{Photon detection}
\label{sec:photon_detection}

The photon detection we consider is a two-outcome measurement process, modeled by two POVM elements as
\begin{equation}
\Pi_1 = \ket{0}\bra{0} \quad \text{and} \quad \Pi_2 = \sum_{n=1}^{\text{dim}(\mathcal{H})} \ket{n}\bra{n} ,
\end{equation}
where $\ket{n}$ denotes the Fock state with $n$ photons and $\dim(\mathcal{H})$ is the dimension of the truncated Hilbert space. The operator $\Pi_1$ is a rank-1 projector corresponding to the outcome \textit{`no photon detected'}, while $\Pi_2$ is a rank-$(\dim(\mathcal{H})-1)$ projector corresponding to the outcome \textit{`photon(s) detected'}. 

\begin{figure}
\centering
\includegraphics[width=\linewidth]{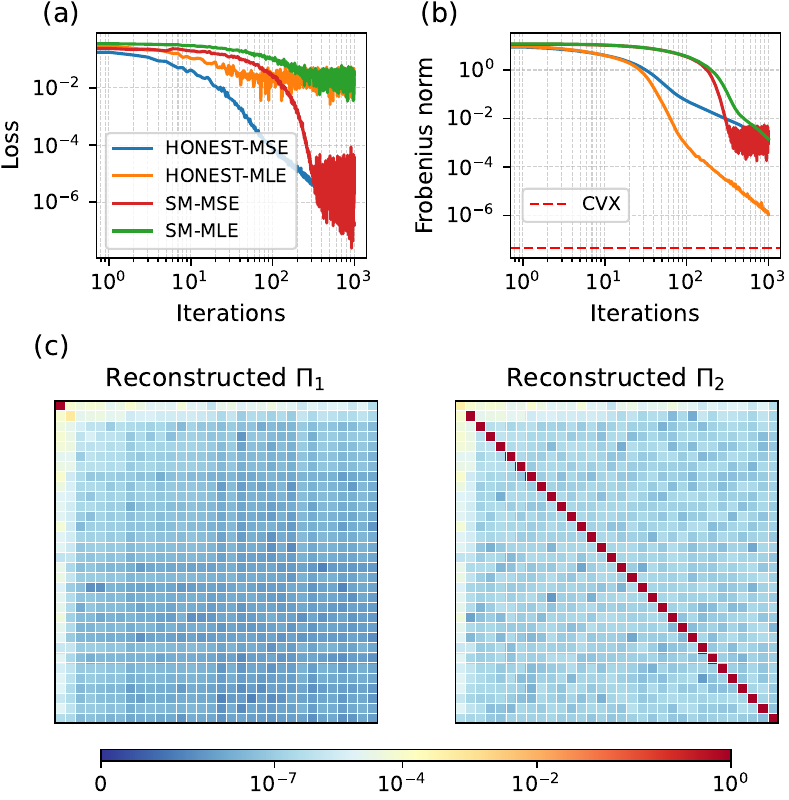}
\caption{Performance of SGD-QMT for photon detection. 
(a) Loss and (b) average Frobenius norm, as functions of the iteration number for HONEST-MSE (teal), HONEST-MLE (orange), SM-MSE (red), and SM-MLE (green). 
(c) The reconstructed POVM elements, $\Pi_{1}$ and $\Pi_{2}$, obtained using HONEST-MLE. Note that the POVM heatmaps are displayed with nonlinear scaling, enhancing the visibility of small entries while compressing larger ones, to highlight fine details in the low-value regions.}
\label{fig:photon_detection}
\end{figure}

In \figref{fig:photon_detection}, we present the results of a numerical study of the performance of our SGD-QMT algorithms for photon detection measurements, using $\alpha = 5$. The first row of the figure shows benchmarking results: the loss as a function of the number of iterations [\figpanel{fig:photon_detection}{a}] and the corresponding average Frobenius norm computed over two POVMs, also as a function of the number of iterations [\figpanel{fig:photon_detection}{b}]. We also plot the Frobenius norm obtained using the CCO-CVX method [dashed red line in \figpanel{fig:photon_detection}{b}] for comparison. The average runtime for completing 1000 iterations is under 3 seconds for all SGD-QMT methods, while CCO-CVX obtained the solution in 19 seconds. The results show that HONEST-MLE (orange curves) exhibits significantly better performance in terms of reconstruction quality than the other three methods [HONEST-MSE (teal), SM-MSE (red), and SM-MLE (green)]. In particular, the Frobenius norm for HONEST-MLE decreases to about $10^{-7}$ within 1000 iterations and continues trending downward, while HONEST-MSE and SM-MLE only reaches $10^{-3}$ in the same amount of iterations (and also continue trending downward) and SM-MSE plateaus around $10^{-3}$.

Figure~\figpanelNoPrefix{fig:photon_detection}{c} displays a tomographic reconstruction of POVM elements obtained using HONEST-MLE: $\Pi_1$ (left) and $\Pi_2$ (right), in the form of heatmaps, each on a $32 \times 32$ dimensional square grid. Both heatmaps show that the reconstructed POVM matrix elements (off-diagonal) are recovered with more than seven-digit accuracy, while the elements (off-diagonal) in the first two rows and columns are accurate to at least four digits. Moreover, the respective dominant diagonal elements (dark red pixels) are consistently recovered with near-perfect accuracy across both POVMs, closely matching the ideal value of one. This result represents excellent agreement with the ideal operators $\Pi_1 = \mathrm{diag}(1,0,\dots,0)$ and $\Pi_2 = \mathrm{diag}(0,1,\dots,1)$. 


\subsubsection{Photon counting}
\label{sec:photon_counting}

Photon counting in CV systems---also known as photon number resolving detectors, is mathematically equivalent to projective measurements in the computational basis in multi-qubit DV systems. The key distinction, however, lies in the choice of probe states, which are fundamentally distinct in the two cases. Photon counting is modeled with $k = \dim(\mathcal{H})$ number of rank-1 POVM elements, each of dimension $k \times k$, defined as
\begin{equation}
\{\Pi_{i} = \ket{i-1}\bra{i-1}\}_{i=1}^{\dim(\mathcal{H})} ,
\label{eq:photon_counting}
\end{equation}
where $\ket{i-1}$ denotes the Fock state with $i-1$ photons. 
%
%

\begin{figure}
\centering
\includegraphics[width=\linewidth]{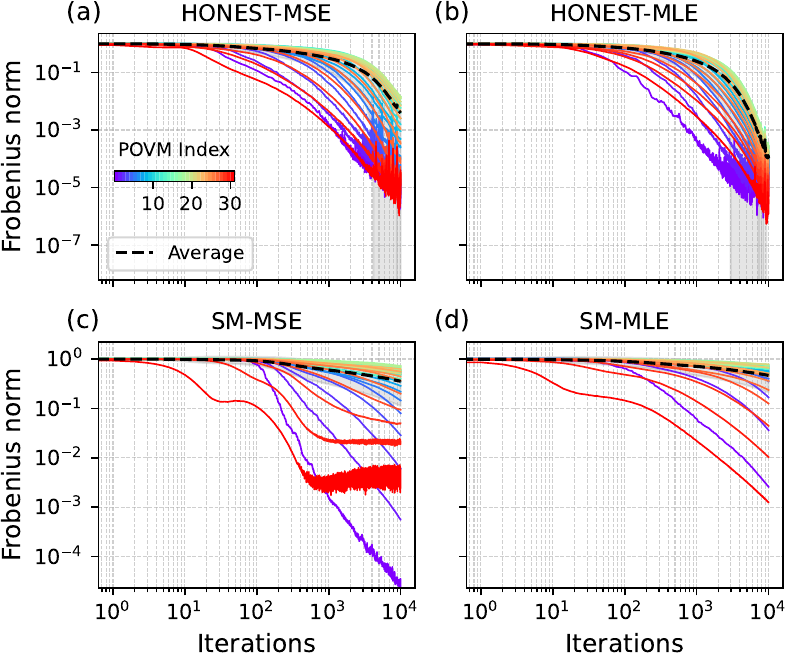}
\caption{Performance of SGD-QMT data-processing algorithms for photon counting. The plots show the Frobenius norm for the reconstructed POVMs as a function of the number of iterations for
(a) HONEST-MSE,
(b) HONEST-MLE,
(c) SM-MSE, and 
(d) SM-MLE.
Each POVM element's Frobenius distance is plotted in a distinct color, with the black dashed line indicating the average Frobenius norm over all 32 POVM elements and the gray shaded region denoting the standard deviation of this average. In all four methods, the total iterations are set to 10000. The average total runtime for all methods is around one minute.}
\label{fig:photon_counting}
\end{figure}

In \figref{fig:photon_counting}, we show results from numerically implementing our SGD-QMT algorithms on photon counting. We set the maximum iteration count to 10000 and used $\alpha = 9$ to reconstruct the 32 POVM elements described in \eqref{eq:photon_counting}, each of dimension $32 \times 32$.

Compared to photon detection (\figref{fig:photon_detection}) and other measurement scenarios we studied in DV systems (Figs.~\ref{fig:time_complexity}, \ref{fig:time_complexity_pauli}, and \ref{fig:iteration}), SGD-QMT of photon counting turns out to be substantially more demanding, in terms of the number of iterations required to achieve a comparable reconstruction quality as measured by the Frobenius norm. Among the SGD-QMT algorithms, HONEST-MLE [\figpanel{fig:photon_counting}{b}] clearly outperforms the alternatives. It achieves an average Frobenius norm as low as $\sim 10^{-4}$ and continues to be monotonically decreasing, whereas HONEST-MSE reaches $\sim 10^{-3}$ with the same number of iterations. The SM parameterizations (SM-MSE and SM-MLE) perform clearly worse; they exhibit very slow convergence rates and their average Frobenius norms do not even approach $10^{-1}$. All four methods took around one minute to finish 10000 iterations (around 160 iterations per second). Note that we have not included CCO-CVX results here because that method failed to produce a solution within an hour, consistent with what we observed for the five-qubit Pauli-projector case in \figref{fig:time_complexity_pauli}.

\begin{figure*}
\centering
\includegraphics[width=\linewidth]{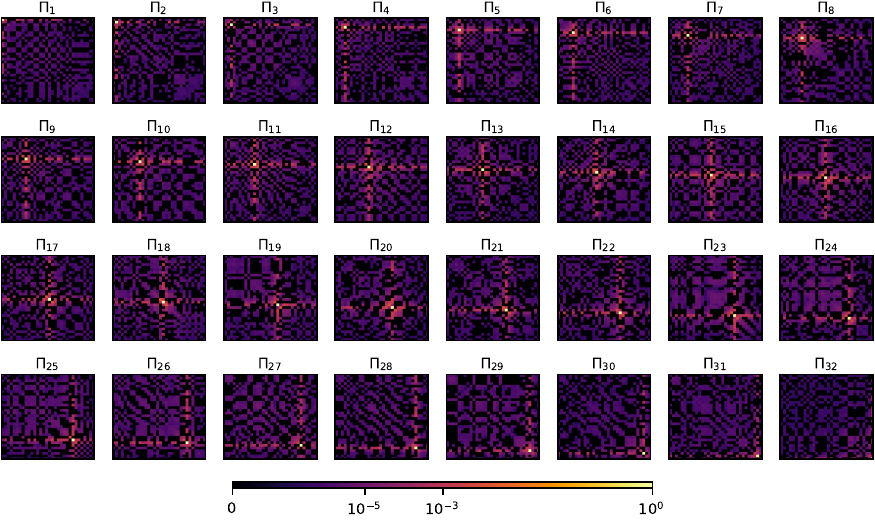}
\caption{Tomographic reconstruction of 32 photon-counting POVM elements using HONEST-MLE, shown as heatmaps. Each heatmap is a $32 \times 32$-dimensional square grid, where the color of each pixel indicates the value of the corresponding matrix element according to the color legend. The heatmaps are displayed with nonlinear scaling to enhance the visibility of small entries while compressing larger ones, thereby highlighting fine details in low-value regions.}
\label{fig:povms_grid_photon_counting}
\end{figure*}

Additionally, \figref{fig:povms_grid_photon_counting} presents the tomographic reconstruction of all 32 POVM operators, shown as heatmaps arranged in an $4 \times 8$ grid. Each heatmap corresponds to a single POVM operator and is represented on a $32 \times 32$ square grid, where each pixel denotes a matrix element and its color encodes the value of that element. These operators were recovered using the HONEST-MLE algorithm, achieving an average Frobenius norm on the order of $10^{-4}$. The heatmaps demonstrate highly accurate recovery of nearly all matrix elements, with precision exceeding five significant digits (dark purple to black regions). Interestingly, for a given POVM operator $\Pi_i$, the elements in the $i$th row and $i$th column exhibit high, but comparatively lower, accuracy than other elements, typically within three to five significant digits (dark purple to dark reddish tones)---a trend also observed in the photon-detection case [see \figpanel{fig:photon_detection}{c}]. Moreover, the dominant $i$th diagonal element (bright yellow pixel) is consistently recovered with near-perfect accuracy across all POVMs, closely matching the ideal value of one.



\section{Conclusion and outlook}
\label{sec:conclusion}

We have developed mini-batch stochastic gradient descent (SGD) data-processing algorithms for quantum measurement tomography (QMT), to extract the POVM elements that characterize an unknown measurement apparatus for quantum systems. Within this framework, we introduced two distinct parameterizations for the POVM elements: (i) the SM parameterization, based on Stiefel-manifold optimization, and (ii) the HONEST parameterization, which utilizes Hermitian operator normalization via an eigenvalue-scaling technique. We showed that both parameterizations inherently satisfy the positivity and normalization constraints required for physically valid POVM elements, and can be seamlessly integrated into the SGD optimization routine to enable valid, accurate, and fast QMT.

We demonstrated the applicability of our SGD-QMT algorithms through numerical simulations on both discrete-variable (DV) and continuous-variable (CV) systems, covering a broad range of measurement processes. In the DV systems, we evaluated the performance of SGD-QMT methods for up to six qubits across three key measurement scenarios: (i) full-rank random POVM sets, (ii) rank-1 Pauli projective measurements, and (iii) rank-1 projective measurements in the computational basis. Similarly, in the CV systems, we focused on two fundamental measurement setups: (i) photon detection and (ii) photon counting. Through simulations we showed that our proposed SGD-QMT algorithms achieve high reconstruction accuracy with substantially reduced computational cost compared to state-of-the-art constrained convex optimization (CCO) using the CVX tool. For example, in the five-qubit case involving 32 full-rank POVMs, SGD-QMT converges in a few tens of seconds, whereas CCO-CVX requires about 1000 seconds ($\sim 15$ minutes) to reach a comparable level of reconstruction quality. All the numerical simulations were carried out on standard laptop with 18 GB of RAM and no dedicated GPU. 

Within the SGD-QMT framework, we examined two types of loss functions: (i) mean squared error (MSE) and (ii) average negative log-likelihood, inspired by maximum likelihood estimation (MLE). To evaluate reconstruction quality, we employed two benchmarking metrics: (i) the Frobenius norm and (ii) the Wasserstein distance. Numerical simulations revealed that all four resulting SGD-QMT variants (two parameterizations combined with two loss functions) share the same time complexity, in terms of iterations per second. However, among these methods, HONEST-MLE consistently outperforms the others (HONEST-MSE, SM-MSE, and SM-MLE), in terms of convergence rate, achieving up to several orders of magnitude higher accuracy in estimating the underlying POVM elements within the same number of iterations (see Figs.~\ref{fig:time_complexity_pauli}, \ref{fig:iteration}, \ref{fig:photon_detection}, and \ref{fig:photon_counting}). 

Our numerical results further highlight that photon counting in CV systems is significantly more computationally demanding than the other measurement processes considered, as reflected in the number of iterations required to achieve high reconstruction accuracy. For instance, even with 10,000 iterations, the best-performing variant, HONEST-MLE, reaches an average Frobenius norm of about $10^{-4}$, whereas in other measurement scenarios, significantly better accuracy is obtained within only 1000--1500 iterations. In particular, five-qubit rank-1 projective measurements (in the computational or Pauli basis) in DV systems are equivalent to photon counting with 32 projectors in CV systems. Yet, despite this equivalence, the performance of SGD-QMT differs substantially between the two scenarios. This finding highlights the critical role of choosing appropriate probe input states. Nonetheless, even at an average Frobenius norm of $10^{-4}$, all matrix elements of the 32 POVM operators characterizing photon counting are reconstructed with very high accuracy, as illustrated in \figref{fig:povms_grid_photon_counting}.

Consequently, the SGD-QMT algorithms developed in this work offer a versatile framework for comprehensive characterization of moderately sized DV and CV systems, covering a broad spectrum of measurement setups relevant to advancing quantum technologies. To facilitate such applications, we have made our SGD-QMT codes freely available here~\cite{sgd-qmt-python}, allowing users to readily apply, adapt, and extend them with additional customizations, as described in \secref{sec:customization}, according to prior knowledge or specific tomography goals.

As an outlook, although our SGD-QMT methods can be applied to reduced data sets, a comprehensive performance study of their data requirements for physically relevant measurement processes---such as Pauli projective measurements in the computational basis, photon detection, and photon counting---will require more extensive numerical simulations. These investigations can be pursued in future work, potentially incorporating the additional modifications discussed in \secref{sec:customization}. Moreover, identifying optimal hyperparameters for a given algorithm and system dimension remains an open and challenging problem. 

Furthermore, the HONEST parameterization can be naturally extended---with only minor modifications---to quantum process tomography (QPT), enabling direct and physically valid reconstruction of sets of Kraus operators. While the SM parameterization has already been applied to QPT~\cite{ahmed-prl-2023}, our numerical simulations demonstrate that the HONEST parameterization consistently outperforms SM for QMT, making it likely the more powerful approach for QPT as well.

Finally, we note that with our SGD-QMT contribution, we complete the gradient-descent based tomography trio---QST, QPT, and QMT. In future work, one could envision combining all three into a single unified optimization problem~\cite{xiao-arxiv-2025}, enabling self-calibrated tomographic characterization of a quantum system, similar in spirit to GST~\cite{xiao-arxiv-2025} but with significantly improved data-processing efficiency.



\begin{acknowledgments}

We acknowledge fruitful discussion with Tangyou Huang. 
This work was supported by the Knut and Alice Wallenberg Foundation through the Wallenberg Centre for Quantum Technology (WACQT), and by the Horizon Europe programme HORIZON-CL4-2022-QUANTUM-01-SGA (project 101113946, OpenSuperQPlus100).
AFK is also supported by the Swedish Foundation for Strategic Research (grant numbers FFL21-0279 and FUS21-0063).

\end{acknowledgments}


\appendix


\section{SGD-QMT with noisy data}
\label{app:noisy data}

In this appendix, we numerically investigate the performance of our SGD-QMT data-processing algorithms in the context of $N$-qubit projective measurements in the computational basis: $\Pi = \{ \ket{0}\bra{0}, \ket{1}\bra{1} \}^{\otimes N}$, evaluated on noisy data sets (see \secref{sec:Results-DV-Projective} for the results obtained with noiseless data). This setting is of particular practical relevance, as projective measurements in the computational basis constitute the primary measurement scheme across most current quantum computing platforms. In realistic experiments, however, the acquired data are inherently noisy, being degraded by decoherence and state-preparation errors.

For demonstration purposes, we prepare noisy input states $\{\tilde{\rho}_j\}$ by applying a depolarizing channel with noise strength $\lambda$ to the ideal input states $\{\rho_j\}$ defined in \eqref{eq:input_states}. The depolarizing channel acts as
\begin{equation}
\tilde{\rho}_j = (1-\lambda)\rho_j + \lambda \frac{I}{2^N},
\end{equation}
where $\tilde{\rho}_j$ denotes the corrupted input state under depolarizing noise and $I$ is the $2^N \times 2^N$ dimensional identity matrix. We then apply our SGD-QMT algorithms to noisy data $\{ \tilde{p}_{ij}\}$ generated as $\tilde{p}_{ij} = \text{Tr}(\Pi_i \tilde{\rho}_j)$.

\begin{figure}
\centering
\includegraphics[width=\linewidth]{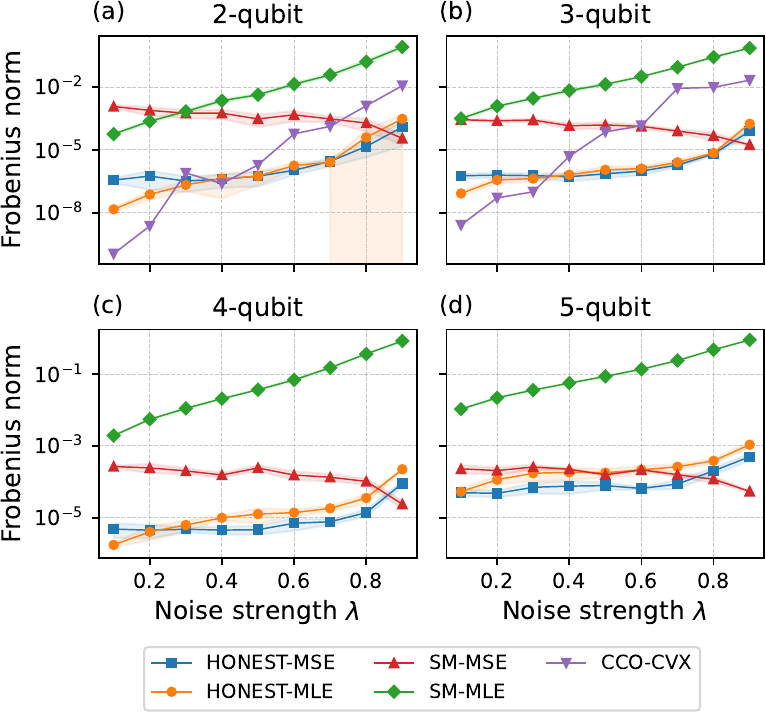}
\caption{Performance of SGD-QMT algorithms for $N$-qubit projective measurements in the computational basis under depolarizing noise, evaluated using the average Frobenius norm ($y$ axes) as a function of depolarizing noise strength $\lambda$ ($x$ axes), for systems with
(a) two, 
(b) three,
(c) four, and
(d) five qubits. 
Results are shown for HONEST-MSE (teal), HONEST-MLE (orange), SM-MSE (red), SM-MLE (green), and CCO-CVX (purple, up to three qubits). The average Frobenius norm is obtained over 15 runs of each method for each given value of $\lambda$. The shaded regions indicate the corresponding standard deviation.}
\label{fig:depo_noise}
\end{figure}

In \figref{fig:depo_noise}, we present a numerical study of the performance of our SGD-QMT algorithms for projective measurements in the computational basis on systems with two to five qubits, and compare it with CCO-CVX (up to three qubits due to time complexity). For small noise-parameter values, CCO-CVX beats SGD-QMT in reconstruction quality, but at high noise values the versions of SGD-QMT using the HONEST parameterization perform better according to that measure. Among the SGD-QMT methods, the results clearly show that the HONEST parameterization consistently outperforms the SM approach. Both HONEST-MSE and HONEST-MLE achieve several orders of magnitude lower Frobenius error across all system sizes and all noise levels except the highest ones, while SM-MLE exhibits the poorest performance in the presence of depolarizing noise. Notably, even at strong depolarizing noise ($\lambda = 0.9$), HONEST-MSE and HONEST-MLE achieve average Frobenius errors as low as $10^{-3}$ (in the five-qubit case; \figpanel{fig:depo_noise}{d}), whereas SM-based methods remain significantly less accurate. Note that, for consistent comparison, the number of iterations was fixed to 1000 across all cases. Although larger systems typically benefit from additional iterations, HONEST already reaches high accuracy within this limited budget. This result highlights not only the efficiency, but also the robustness of the HONEST parameterization under realistic noise conditions.


\section{Parameter-update rules and hyperparameters in SGD-QMT algorithms}
\label{app:hyperparameters}

We adopt different gradient-based methods depending on the parameterization, as explained in \secref{sec:GD-Algorithms}. Specifically, for the SM parameterization we use vanilla gradient descent [see \eqref{eq:sm_update}], while for the HONEST parametrization we employ the \textit{Adam} optimizer. The Adam update rule can be briefly expressed as~\cite{ruder2016overview}
\begin{equation}
\bm{\theta}_{t} \gets \bm{\theta}_{t-1} - \eta \cdot \bm{\hat{m}}_{t} /(\sqrt{\bm{\hat{v}}_{t}} + \epsilon) ,
\label{adam_gd}
\end{equation}
where $\bm{\theta}_{t}$ is the parameter vector, which in the HONEST parameterization is a list of complex matrices $\{ T_i \in \mathbb{C}^{d \times d} \}$, at $t$th step. The vectors $\bm{\hat{m}}_{t}$ and $\bm{\hat{v}}_{t}$ represent bias-corrected first and second moments, respectively. We fix $\epsilon = 10^{-8}$ to avoid divergence. The parameter $\eta$ denotes learning rate or step size. The detailed pseudo-code for the Adam algorithm can be found in Ref.~\cite{gaikwad-arxiv-2025}. 

Our SGD-QMT algorithms rely on four key hyperparameters: (i) state-batch size $m$, (ii) POVM-batch size $n$, (iii) learning rate $\eta$, and (iv) decay factor $\alpha$. At the $t$th iteration, the mini-batch is constructed by randomly sampling $m$ input states $\{\rho_i\}_t^{(m)}$ and $n$ POVM operators $\{\Pi_i\}_t^{(n)}$, resulting in an effective mini-batch size of $m \cdot n$, as specified in \eqref{eq:gradient}. In all simulations presented in this work, the POVM-batch size $n$ is chosen to equal the total number of POVMs, while the state-batch size $m$ is fixed at 50 for all methods and scenarios (with the exception of the two-qubit case, since the maximum state-batch size there is 16). The learning rate $\eta$ is set to 0.01 for the HONEST parameterization and 0.05 for SM. Note that the decay parameter $\alpha$ is required only for the SM parameterization under the vanilla gradient descent update rule [see \eqref{eq:sm_update}], reducing the learning rate as $\eta \cdot \alpha$ at each iteration to suppress fluctuations near convergence; it is fixed to 0.99 across all scenarios. 



\bibliography{References}

\end{document}